\documentclass[preprint,aps]{revtex4}
\usepackage{epsfig}
\usepackage{amsmath}
\usepackage{graphicx}
\usepackage{dcolumn}
\usepackage{bm}
\usepackage{color}

\begin{document}

\title{A novel Topological Model for Nonlinear Analysis and Prediction for Observations with Recurring Patterns}
\author{Sajini Anand P S}
\affiliation {International Centre for Theoretical Sciences, ICTS-TIFR, Bangalore 560089. }
\email{sajini.anand@icts.res.in}
\author{Prabhakar G Vaidya}

\email{pgvaidya@gmail.com}
\date{\today}

\begin{abstract}

The paper introduces Atlas for Recurrence neighborhood~(ARN) Model-- a novel topological method for prediction and modeling for a nonlinear time-series that exhibits recurring patterns. According to the model, global manifold of a reconstructed state-space can be approximated by a few overlapping recurrence neighborhoods. The inherent redundancy structure of the delay-embedding procedure and the property of recurrence are used to reduce the computational load, which is inevitable in nonlinear analysis. The modeling and prediction capabilities of the model are demonstrated using (i) a numerical data generated by a dynamical system: the Duffing oscillator and (ii) a real-world data: an Electrocardiogram ECG recording of a healthy human. A potential application of the proposed model is demonstrated for a multivariate cardiovascular data set that exhibit recurring patterns. Real-time monitoring of cardiovascular signals are essential in clinical research and corruption of data are very common. It is a challenging task for a model to perform cognitive functions of brain such as predicting loss of data or identifying noises in the physiological data. Potentials of the proposed ARN model to perform a cognitive task--`the prediction of gaps or loss of data based on the contextual information' are explored in the study reported here, analysing a set of $100$ multichannel cardiovascular recordings. 

\textbf{One Sentence Summary:} A topological model for prediction and modeling for a time-series that exhibits recurring patterns.
\end{abstract}

                              
\maketitle

\section{Introduction}

Chaos theory and topology in the past few decades have given many useful insights to understand data generated by nonlinear systems. In 1980, Packard \emph{et al.} observed that the state-space of a nonlinear system could be reconstructed from a time-series generated by that system~\cite{packard1980}. It opened up a possibility of associating geometrical and topological structures with observed data~\cite{packard1980,muldoon1993topology}. Later Takens and Ma$\tilde{n}\acute{e}$ gave a proof and listed conditions to be met for any reconstruction to be diffeomorphic to the original state-space~\cite{takens1981,mane1981dynamical}. State-space reconstruction techniques are used to develop a set of equations or maps for a system, from an observed time-series for prediction or description of the dynamics~\cite{crutchfield1987,breeden1990,aguirre2009}. The standard methods broadly fall into two categories: (i) Global methods that find equations valid for the entire state-space~\cite{crutchfield1987,cremers1987} and (ii) Atlas methods that develop local charts/atlas for small neighborhoods of the state-space~\cite{farmer1987,farmer1988,sajini2014}.

\vspace{0.1in}

Nonlinear signals often mimic random signals despite their deterministic origin as are non-periodic and non-stationary. Recurrence plots and Poincare Sections are some classic techniques used for nonlinear analysis~\cite{marwan2008,ott2002,alligood1996chaos}. Recurrence~(the property that a typical trajectory of the system keeps on visiting the neighborhood of a particular state in the state-space) is a characteristic of nonlinear chaotic systems and identification of recurrence patterns is a prominent method for nonlinear data analysis~\cite{zbilut1992,webber2005recurrence}. This paper introduces a novel topological model based on atlas for analyzing a specific class of nonlinear time-series that exhibit the property of recurrence.

\vspace{0.1in}

The problem of state-space reconstruction is defined as follows. Assume that some discrete measurements or observations $S_i \in R $ of a dynamical system $ \dot X = F(X)$ where $X \in R^{d}$ and $S = H(X)$ are available in a finite time- interval $(0,\Gamma)$. Given that the evolution function $F$, observation function $H$ and the dimension of the system $d$ are unknown, what are the possibilities of identifying the properties of the system for prediction or analysis, solely based on a finite time-series $\{ S_i \}$ generated by it?
\vspace{0.1in}

Delay, derivative and filtered data-coordinate embeddings are standard methods that address this question. They build a global multidimensional state-space using either delay vectors~\cite{takens1981,mane1981dynamical}, derivatives~\cite{aguirre2009,gouesbet1994global} or filtered data vectors~\cite{Broomhead1986,broomhead1992} generated from observations~\cite{abarbanel2012analysis}. Embedding methods assume that the data used for reconstruction is an `observable' \emph{i.e.,} it contains information about all state-variables of the system. In principle, we can determine the observability of any observations iff the system equations are known~\cite{Robenack_observability,letellier2005observability}. When the system equations are unknown, which is most of the reality, it is still ambiguous how to determine the observability of any observation. In this scenario, validation of a model through state-space reconstruction relies on the availability of multiple observations of the system, existence of many possible models and comparisons of the derived models for further analytical investigations~\cite{mangiarotti2012polynomial, letellier2009}.
\vspace{0.1in}

While global models look for equations for the entire reconstructed state-space~\cite{mangiarotti2012polynomial}, atlas models partitions the reconstructed space into neighborhoods within which the dynamics can be approximated by local constant predictors~\cite{farmer1987, kennel1992}, weighted predictors or other linear polynomials~\cite{linsay1991efficient}.  These models are known to use a lot of CPU resources in searching for nearest neighborhoods in the embedded space. The demands for resources are directly proportional to the required embedding dimension~\cite{grassberger1991nonlinear}. The model proposed here drastically reduces the computational time for searching neighborhoods by utilizing the recurrence property of the time-series and the redundancy of delay-coordinate embedding.
\vspace{0.1in}

In what follows, Section II revises delay-coordinate embedding as a tool for modeling a system based on a time-series generated by it.  Section III proposes a topological model 'Atlas for Recurrence Neighborhood~(ARN)' for a time-series that exhibits recurring patterns; the model is based on  a generalization of Eckmann's Recurrence-plots, for identifying a recurring pattern in the time-series and its neighborhood of recurrence. Section IV details a parametrization scheme and the generation of equations for evolution and conjugacy maps across the neighborhoods. Section V presents the results of data analysis for a numerically generated data of Duffing oscillator under chaos. Section VI discusses the application of the method for a real-world data: a univariate ECG data of a healthy human. Section VII explores an application of the ARN model for prediction, for a multivariate cardiovascular data set. Concluding discussions are presented in Section VII.
\vspace{0.1in}

\section{Delay-coordinate Embedding as a Modeling tool}

Delay-coordinate embedding is a convenient~(not unique) representation of the state-space using delay-coordinates generated from the data. The scheme of state-space reconstruction rely on an assumption of observability of the time-series~(the observation being mathematically generic monitoring all degrees of freedom of the system, thus containing information about other state-variables). In principle, observability of any observation can be verified if generating equations are known~\cite{letellier2005observability}. However, an inability to meet the requirement of observability with acquired data in practice can affect the faithfulness of the reconstruction.
\vspace{0.1in}

Assume that $S_i \in R $ are observations of a dynamical system $(X, F)$ where $X \in R^{d}$  and $\dot X = F(X)$  such that, $S_i= H(X_i)$ where $H : R^P \mapsto R $ denotes an observation function. The embedding theorems establishes the existence of a faithful reconstruction of the original state-space in some $R^P$ (where $P > 2d$) using P-dimensional delay vectors of the form $ \{ S_n , S_{n+ \tau} , \ldots , S_{n+(P-1)\tau }\}$ made from observations, where $P$ is the embedding dimension and $\tau$ is the time-delay. Thus, the delay map $D: R^d \mapsto R^P$ is generically faithful to the original system state-space if the embedding dimension $P> 2d$, the true dimension of the system~\cite{takens1981,mane1981dynamical}. The theorem also assumes the availability of infinitely many observations in a perfect noise-free scenario. However, in practice, we deal with a finite amount of discrete data in a highly probable noisy scenario. In addition to that, we need to face more limiting factors such as, the system noise along with the difference in time-scales between different parts of the system and its effects on observations~\cite{ott1994coping}.
\vspace{0.1in}

Many heuristic methods are available to estimate the embedding dimension $P$ and time-delay $\tau$ from data, since the evolution function $F$ and the dimension $d$ of the system under study are unknown, in most of the reality~\cite{bradley2015CHAOS}. One general approach for estimating embedding dimension from data is to increase the dimension until all false near neighbors~(FNN) reduce to zero~(if data is noise free) or to a minimum~(if data is noisy). In general, our faith in false near neighbor~(FNN) tests are inversely related to the noise present in the data~\cite{kennel1992FNN}. FNN-estimates can be considered as a lower bound for the embedding dimension for reconstruction, as the attractor can be assumed to be unfolded by then~\cite{abarbanel2012analysis}. Takens theorem does not specify an upper bound on embedding dimension, and using a dimensional estimate much higher than the actual value is a general practice to deal with under-sampling and noise in data. In general, one would prefer a lower dimensional model as it is economical regarding data compared to its high dimensional counterpart. Also, spurious instabilities are reported in global analysis when a system is embedded in a dimension higher than what is required~\cite{bezruchko2001}.
\vspace{0.1in}

The model proposed here uses a high estimate for embedding dimension $P$ which is higher than that is suggested in literature~\cite{abarbanel2012analysis}, as aim of the model is to identify a neighborhood for a recurring pattern~(of length $P$) in a time-series using delay-coordinate embedding in $R^P$.
It uses an atlas-based procedure for partitioning the reconstructed space into some finite number of neighborhoods and approximating the dynamics in the neighborhoods using various maps~\cite{farmer1987, kennel1992,linsay1991efficient}. We use a data-driven parametrization scheme for manifold learning by mapping the data from a high dimensional delay embedding space to a low dimensional parametric space. A recurrence neighborhood for the specific pattern is parameterized in $R^d$. We assume that neighbourhood lives in a thinner manifold of dimension $d$ even though the data belongs to high dimensional space $R^P$ since $d<<P$. Section V and VI deal with highly sampled pattern~($800$ data points), hence $P$ used is $800$, and section VII has under-sampled pattern~($110$ data points) in real-world data, hence $P$ used is $110$.

\vspace{0.1in}

Time-delay $\tau$ can be estimated based on available rules of thumb~\cite{abarbanel2012analysis,bradley2015CHAOS}. The process of delay-coordinate embedding requires an identification of a basis to unfold the attractor. One would prefer a set of basis vectors that are possibly independent of each other for this purpose. Since delay vectors by their structure carry much mutual information across them, the value of the smallest time-lag for which the mutual information is a minimum across the delay vectors is proposed as a reasonable delay for embedding~\cite{fraser1986}. However, in practice when the data is noisy, and possibly under-sampled, the mutual information curve may not show a minimum, so a delay of unity ($\tau= 1$) is suggested~\cite{abarbanel2012analysis}. We stick to the delay of unity as the proposed model is designed to benefit from the redundancy of information that could be available in the delay-vectors. Sampling frequency was known for the real-world data we considered for analysis. We set delay as unity for delay-embedding for two specific reasons (i) signals recorded were human physiological measurements which could be under-sampled and noisy  (ii) the proposed method wants to exploits whatever mutual information that is present in the delay vectors. The redundancy of information between delay vectors is a significant factor that reduces the computational cost of data analysis.

\vspace{0.1in}

\section{Neighborhood of Recurrence in Delay-coordinate Embedding}

Recurrence was formally introduced in 1890 by Henry poincar$\acute{e}$ as a property of the system to recur as infinitely many times as close to an initial state~\cite{poincare1890}. Later in 1987, Eckmann \textit{et al.} introduced Recurrence Plots to visualize the property of recurrence from a time-series of the system. A system is defined as recurrent if a time-series generated by it keeps on visiting a specific neighborhood of the manifold in the state-space~\cite{eckmann1987}. Following this, analyzing recurrence patterns and finding them in a time-series is a prominent method for data analysis~\cite{zbilut1992,webber2005recurrence,marwan2007} in numerous fields~\cite{webber1994dynamical,wyner1998,antoniou2000,thomasson2001,marwan2008}.

\subsection{Eckmann's Recurrence Plots (RP) for visualization of recurrence}

Recurrence plot (RP) provides a two-dimensional visual representation of a trajectory generated by a dynamical system. Consider a finite time-series $ \{x_n\}_{1 =1 \ldots N}$ that represents a trajectory of the system. Then RP is defined by a 2D matrix as follows:

\begin{equation}
R(i,j) =
  \begin{cases}
    1   & \quad \text {if } {x_i \approx x_j} \\
    0   & \quad \text{ otherwise}\\
  \end{cases}
\end{equation}

when $R(i,j)$ is true,  $(i,j)$ are the pair of times at which the trajectory is in the same $\epsilon$-neighborhood of the state-space, i.e.,\
$(x_i \approx x_j )\implies$ $(x_i - x_j) \leq \epsilon$. Visualization of RP, due to its rich geometrical structure is used as a tool to find the correlations that are present in different scales within the data. Further, a series of measures listed in Recurrence Quantification Analysis (RQA) can be used to define RP quantitatively~\cite{zbilut1992,marwan2008}.

\subsection{Generalized Recurrence (GR) for identifying recurring patterns}

\begin{figure}
\begin{centering}
\includegraphics [width=1 \textwidth]{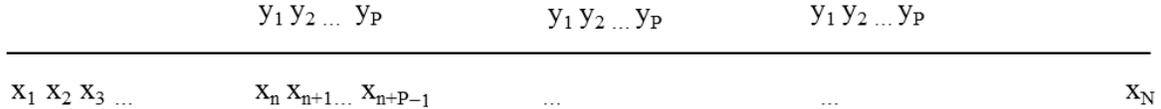}
\caption[]{The schematic depicts the recurrence of a finite sequence $ \{y_p\}_{p =1 \ldots P }$ in a given time-series $ \{x_n\}_{n =1.. N}$ such that $P \ll N$.}
\label{Fig:rec_schematic}       
\end{centering}
\end{figure}

In this section, we define a generalization for RP, to identify recurrence of a single chosen pattern in a time-series as follows.
Let $ \{y_p\}_{p =1 \ldots P }$ be a finite sequence of length $P$, that represents a specific pattern in the given time-series $ \{x_n\}_{n =1.. N}$ such that $P \ll N$ as depicted in Figure~\ref{Fig:rec_schematic}. Refer as an example, Figure~\ref{Fig:5ECGtimeseries}(a) is a specific pattern of an Electro cardiogram (ECG) record that  represents a full cardiac cycle in a finite time-series of ECG signal shown in \ref{Fig:5ECGtimeseries}(b).
In order to identify the sequence $ \{y_p\}_{p =1 \ldots P }$ in the given series, construct a set of $P$ dimensional delay vectors $Z_i$ from $ \{x_n\}_{n =1.. N}$ such that $Z_1 =(x_1,x_2,\ldots x_P)^T$, $Z_2 =(x_2,x_3,\ldots x_{P+1})^T$, $ \ldots Z_n =(x_n,x_{n+1},\ldots x_{n+P+1})^T$. This process is equivalent to  the delay embedding of $ \{x_n\}$ in $ R ^P$ with a delay $1$ and dimension $P$, as $Z\in R^P$~\cite{abarbanel2012analysis}. 

Then the Generalized Recurrence function is defined as follows:

\begin{equation}
 GR(i,j) =
  \begin{cases}
    1   & \quad \text {if } {Z_i \approx Z_j} \\
    0   & \quad \text{ otherwise}\\
  \end{cases}
   \end{equation}

when $GR(i,j)$ is True, $(i,j)$ are the pair of times at which the delay vectors $Z_i, Z_j$ are significantly close to each other.
Alternately, the points $Z_i, Z_j$ of the trajectory of $Z$ (evolution of $P$-dimensional delay vector) are in the some $\epsilon$-neighborhood of the reconstructed state-space in ($R^P$). Here
$(Z_i \approx Z_j )\implies \|(Z_i - Z_j)\| \leq \epsilon$, where $\|(Z_i - Z_j)\| $ represents an appropriate choice of norm of the difference vector in $R^P$. Note that $ GR(i,j)$ is a function of $P$ and it can capture recurrence of all possible sequences of length $P$ in the given time-series $\{x\}$.

\subsection{Neighborhood of Recurrence for a Specific Pattern}

We have seen that $GR(i,j)$ by definition can capture recurrence
of all possible sequences of length $P$ in the given time-series $\{x\}$.
Let us focus on the recurrence of a specific pattern $\{y_p\}$ of length
 $P$, this pattern will be referred as $Z_{ref}$ henceforth.
 Define a similarity measure L for the reference pattern $Z_{ref}$ with
 respect to all other $Z_i \in R^P$, such that
 \begin{equation}
 L_{Z_{ref}}(i)= \|(Z_i - Z_{ref})\|
 \end{equation}
$L$ reflects the recurrence of $Z_{ref}$ in the time-series $\{x\}$,
since $L(i)$ is the distance of $Z_{ref}$ to every other vector $Z_i \in R^P$.
\vspace{0.1in}

Once a specific pattern ($Z_{ref}$ of length $P$) is identified in a time-series, identification of recurrence neighborhood for $Z_{ref}$ is equivalent to finding a neighborhood of the pattern $Z_{ref}$ in a high dimensional delay coordinate embedding space $R^P$.  Visualization of $L$ reveals more properties about the recurrence of $Z_{ref}$ in the time-series:  (i) how frequently the pattern gets repeated in the given time-series, and (ii) the qualitative and quantitative aspect of the recurrence cycle to infer the evolution of trajectory in a reconstructed state-space.
Referring back to the ECG data in Figure~\ref{Fig:5ECGtimeseries}, chosen pattern $Z_{ref}$ is depicted in \ref{Fig:5ECGtimeseries}(a) which is highlighted in the ECG time-series in ~\ref{Fig:5ECGtimeseries}(b) and ~\ref{Fig:5ECGtimeseries}(c) depicts $L(i)$-- the L2 norm of the difference of $Z_{ref}$ to every other vector $Z_i \in R^P$. Note that Figure~\ref{Fig:5ECGtimeseries}(c) effectively reflects the recurrence of the chosen pattern in the series.

\vspace{0.1in}
In order to collect points in a neighborhood of recurrence, one can follow this algorithm: Given a pattern of interest $Z_{ref}$, visualize the similarity measure of $Z_{ref}$ defined in Eq~3. $L$-plot can be used to infer qualitatively (i) the recurrence of $Z_{ref}$ in the time-series and (ii) qualitatively infer the length of consecutive recurrence cycles. Define a threshold $\delta$ to include the all the desired recurrent cycles.  Collect all points that correspond to the local minima of every recurrence cycle as the members of the Recurrence Neighborhood (RN) $ \in R^P$ and index the timings of the recurrence into an array Recurrence Timings (RT) and calculate the length of each recurrence cycle $\tau$ by taking the time-delay between consecutive recurrences.
\vspace{0.1in}

\section{Atlas for Recurrence Neighborhoods (ARN) Model and Parametrization Scheme}

Atlas methods focus on partitioning the embedded space into some finite number of neighborhoods
$\{U_m\}_{m=1}^M $ and then approximate the dynamics in the neighborhoods using various maps
~\cite{farmer1987, kennel1992,linsay1991efficient}. A practical difficulty of this approach
is the computational cost for finding close neighbors that
grows exponentially with embedding dimension~\cite{grassberger1991nonlinear}.
Here we introduce an `Atlas for Recurrence Neighborhood (ARN)' model
specifically for analyzing a time-series that exhibit the property of recurrence.
The redundancy structure of delay embedding along with the recurrence in time-series reduce the computational load that is required otherwise.
The ARN model computes a set of neighborhood preserving
embeddings for Recurrence Neighborhoods reconstructed from the data.
\vspace{0.1in}

\subsection{Parametrization for the Recurrence Neighborhood in $R^d$ where $d<<P$ }

The ARN model is based on a data-driven parametrization scheme. It does manifold learning
by mapping the data from a high dimensional delay embedding space to a low dimensional parametric space following an empirical procedure.
We have seen how to construct a recurrence neighborhood RN for a specific pattern ($Z_{ref}$ of length $P$) in $R^P$ in section III C.
Once RN is identified in $R^P$, it is projected to $R^d$ (where $d<<P$) using a linear transformation $ A : R^N \rightarrow R^d$ described in Eq.~4. Since $d<<P$, we assume that RN lives in a thinner manifold of dimension $d$ even though the data belongs to high dimensional space $R^P$. ARN model is used to approximate the dynamics on the manifold using lower order polynomial maps.
\vspace{0.1in}

The linear transformation A (of dimension $d \times P$ operates on an $P \times 1$ vector of $R^P$ to generate a $d \times 1$ vector of $R^d$) represents the projection: $R^P \rightarrow R^d$.

\begin{equation}
\begin{array}{l}
\displaystyle A_{0,p}=\frac{1}{P} \text{Re} \left( \mathrm{e}^{ \left(\frac{-i 2 \pi p}{h_0}\right )}\right )\\
\displaystyle A_{1,p}=\frac{1}{P} \text{Im} \left( \mathrm{e}^{ \left(\frac{-i 2 \pi p}{h_0}\right )}\right )\\
\displaystyle A_{2,p}=\frac{1}{P} \text{Re} \left( \mathrm{e}^{ \left(\frac{-i 2 \pi p}{2 h_0}\right )}\right )\\
\displaystyle A_{3,p}=\frac{1}{P} \text{Im} \left( \mathrm{e}^{ \left(\frac{-i 2 \pi p}{2 h_0}\right )}\right )\\
\end{array}
\end{equation}

\textbf{Parameter values for Numerical Analysis:} The parameter values for $h_0, P, d$ of $A$ matrix are derived from the data. $P$ is chosen as the total number of points in the pattern of interest $Z_{ref}$, $h_0$ matches with the total number of points in the specific pattern or subpattern of $Z_{ref}$ and the parametric space dimension $d$ is fine-tuned in the range $\{2, 3, 4\}$ based on the efficiency of prediction.

\subsection{Conjugacy between the neighborhoods in $R^N$ and $R^d$}

We have seen in the previous section how the neighborhood RN in $R^P$ is projected to $R^d$ (where $d<<P$) using the linear transformation $ A : R^N \rightarrow R^d$ described in Eq.~4. To demonstrate conjugacy between the neighborhoods in $R^P$ and $R^d$, we need a mapping from $R^d$ to $R^P$ across the neighborhoods.  Mapping from a lower dimension to higher dimension is not straightforward especially if the dynamics is nonlinear. In this case, embedding dimension was set very high ($P\gg d$) to generate a recurrence neighborhood in $R^P$. So we make an assumption that the recurrence neighborhood lies locally on a d-dimensional manifold embedded in $R^P$, though the global manifold of RN lives in $R^P$. This section explains a method to construct a neighborhood preserving map between the embedding space $R^P$ and the parametric space $R^d$ to establish a conjugacy between the manifolds in both the spaces.
\vspace{0.1in}

Let matrices $RX$ and $RY$ represents the recurrence neighborhoods in $R^P$ and $R^d$ respectively. Assume that the neighborhood contains $q$ vectors each. $RX$ has dimension $q \times P$ and $RY$ has dimension $q \times d$. Let $rX$ and $rY$ represent the recurrent neighborhood matrices translated to the origin (the centroid of the neighborhoods is subtracted from neighbors to translate the neighborhood to the origin).  Record the centroid of $RX$ and $RY$ as $\overline{RX}$ and $\overline{RY}$ respectively.
\begin{eqnarray}
\overline{RX}_j = \frac {1}{q}\sum _{i=1}^q{RX_{q,j}}\; \; \;  \; \; \;\mathrm{for \; \; j =1,2 \ldots P}\\
\overline{RY}_j = \frac {1}{q}\sum _{i=1}^q{RY_{q,j}}\; \; \;  \; \; \;\mathrm{for \; \; j =1,2 \ldots d}
\end{eqnarray}
\vspace{0.1in}

Neighborhoods translated to the origin are:
\begin{eqnarray}
rX_{q,j} = RX_{q,j}- \overline{RX}_j \; \; \;  \; \; \;\mathrm{for \; \; all \; \;q}, \; \; \;\mathrm{for \; \; j =1,2 \ldots P}\\
rY_{q,j} = RY_{q,j}- \overline{RY}_j \; \; \;  \; \; \;\mathrm{for \; \; all \; \;q}, \; \; \;\mathrm{for \; \; j =1,2\ldots d}
\end{eqnarray}

Define a conjugacy map $T$, across the neighborhood matrices $rX$ and $rY$ such that,
\begin{equation}
(rY). T = rX
\end{equation}

Now $rX$ of dimension $q \times P$, $rY$ of dimension $q \times d$ are known matrices; $T$ of dimension $d \times N$ can be calculated by pre-multiplying $rX$ with the generalized inverse of $rY$.

\begin{equation}
T  = (rY)^{-1}. rX \\
\end{equation}

$T$ represents a linear map across the neighborhoods $rY, rX$.
Hence, for every vector $\alpha$ in $RY$ there exists a vector $\beta$ in $RX$ such that,
\begin{equation}
\beta= T \alpha + \overline{RX}
\end{equation}

Once $T$ and the centroid $\overline{RX}$ are known, a vector in $R^N$ can be generated from a vector in $R^d$. The transformation $T$ represents the mapping from  $R^d \rightarrow R^N$ as any of the recurrence vectors in $R^N$ can be predicted using their $R^d$ counterparts.

$T: R^d \rightarrow R^N$ of Eq.~11 together with $A: R^N \rightarrow R^d$ of Eq.~6  represents the topological conjugacy across the recurring neighborhoods in $R^{N}$ and $R^d$.

\subsection{Evolution of Dynamics in $R^d$ using Equivalence Classes}

The members of the recurrence neighborhood in $R^d$ can be further classified into different equivalence classes based on their recurrence cycle~$\tau$ (find $\tau$ as suggested in section III C). Collect all the vectors that recur at the same time to belong to one equivalence class.  Evolution maps for a particular equivalence class can be defined as follows.
\vspace{0.1in}

Collect all beginning and end vectors of a specific recurrence cycle ~$\tau$ in $R^d$ to matrices $SX$, $SY$ respectively. Assume $SX$, $SY$ contain q vectors each. Let $sX$ and $sY$ represents the equivalence classes translated to the origin (the centroid of $SX$, $SY$ subtracted from the neighbors to translate to the origin). Record the centroid of $SX$ and $SY$ as $\overline{SX}$ and $\overline{SY}$ respectively. let

\begin{eqnarray}
\overline{SX}_j = \frac {1}{q}\sum _{i=1}^q{SX_{q,j}}\; \; \;  \; \; \;\mathrm{for \; \; j =1,2 \ldots d}\\
\overline{SY}_j = \frac {1}{q}\sum _{i=1}^q{SY_{q,j}}\; \; \;  \; \; \;\mathrm{for \; \; j =1,2 \ldots d}
\end{eqnarray}

Neighborhoods translated to the origin are:
\begin{eqnarray}
sX_{q,j} = SX_{q,j}- \overline{SX}_j \; \; \;  \; \; \;\mathrm{for \; \; all \; \;q}, \; \; \;\mathrm{for \; \; j =1,2 \ldots d}\\
sY_{q,j} = SY_{q,j}- \overline{SY}_j \; \; \;  \; \; \;\mathrm{for \; \; all \; \;q}, \; \; \;\mathrm{for \; \; j =1,2\ldots d}
\end{eqnarray}

Define a transformation V, across the matrices $sX$ and $sY$ such that;
\begin{equation}
 sX. V  = sY \\
\end{equation}

If the specific equivalence class contains $q$ members, $sX$, $sY$ both of dimension $q \times d$ are known matrices; $V$ of dimension $d \times d$ can be calculated by pre-multiplying $sY$ with the generalized inverse of $sX$.
\begin{equation}
V  = (sX)^{-1}. sY \\
\end{equation}

The transformation $V$ is a function of the recurrence cycle $\tau$ associated with the chosen equivalence class.
Once the $V$ matrix is estimated, any random vector $\alpha \in SX$ can be evolved to $\beta \in SY$.

\begin{eqnarray}
\alpha- \overline{SX}& = &r\alpha\\
r\alpha. V & = &r\beta \\
r\beta + \overline{SY} & =& \beta
\end{eqnarray}

$\beta$ represents the evolution of $\alpha$ at the end of the recurrence cycle $\tau$ in space $R^d$. In this manner, one can evolve a vector in $R^d$ using the evolution maps for all the equivalence classes (for all recurrence cycles) along with corresponding centroids.

\subsection{Prediction and Modeling using ARN model}

A significant share of complexity in prediction for the nonlinear time-series is attributed to the the dynamics that is responsible for the variability in recurrences of the chosen pattern. Procedure for ARN model is as follows. Analyze time-series to find the recurrence cycles of the chosen pattern. Once the recurrence timings RT and cycles $\tau$ are recorded, prediction can be done based on 2 types of maps: (i) the set of evolution maps (specific to the equivalence classes and recurrence cycle) in a low dimensional space $R^d$ and (ii) the conjugacy maps that exist across the low and high dimensional spaces ($R^d \rightarrow R^P$). Given an initial vector and its recurrence cycles, (i) evolving it in $R^d$ according to the recurrence cycles,(ii) finding the corresponding patterns in $R^P$ and (iii) stitching the predictions at appropriately according to the recurrence timings RT will complete the prediction procedure for ARN model.

Following sections explain the capabilities of ARN model for the purpose of modeling and prediction in different data analysis scenarios: Section V-- numerically generated data by Duffing oscillator under chaos, Section VI-- a highly sampled and lengthy ECG record of a healthy human heart and Section VII-- under-sampled, noisy and short data segments of  multichannel cardiovascular signals recorded from patients in ICU settings.  These sections demonstrate how beneficial and robust the predictions are for the proposed model.

\section{\label{sec:level1}Analysis of Data: Duffing oscillator under chaos }
\begin{figure}
  \begin{center}
    \includegraphics [width= 0.5 \textwidth]{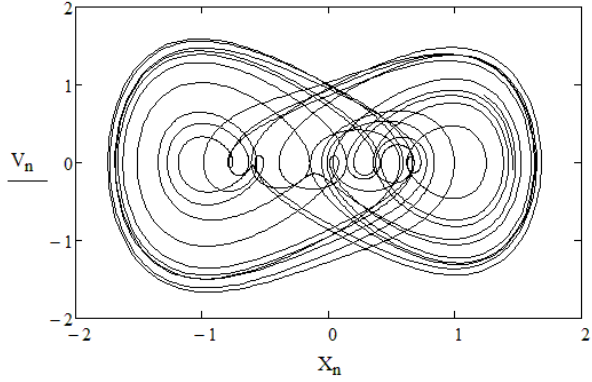}
    \caption{Numerically generated state-space of the Duffing oscillator under Chaos by integrating Eq.~21, $X_n, V_n$ being numerical values for $x, \dot{x}$} \label{Fig:SS}
  \end{center}
\end{figure}

This section elaborates the proposed method of modeling using a time-series numerically generated by Duffing oscillator. Duffing Oscillator is a periodically forced system with a nonlinear elasticity as represented by a second-order differential equation given in Eq.~21. The model is found to be robust for a wide range of parameters-- from periodic motion to chaos. Figures presented in this section are for the following parameter values of Eq.~21 ($c=0.04496$, $k=0$, $\rho=1$, $\alpha=0$, $F=1.03$) for which the system exhibit chaos. Figure~\ref{Fig:SS} represents a numerically generated state-space of the oscillator for a time-series generated from an initial condition $(1.226, 0.868)$.
\begin{equation}
\ddot{x}+ c \dot{x}+k x +\rho x^3 = F cos(\omega t+ \alpha)
\end{equation}

\subsection{ Identification of a Pattern and its Recurrence Neighborhood}

\begin{figure}
  \begin{center}
    \includegraphics [width= 1 \textwidth]{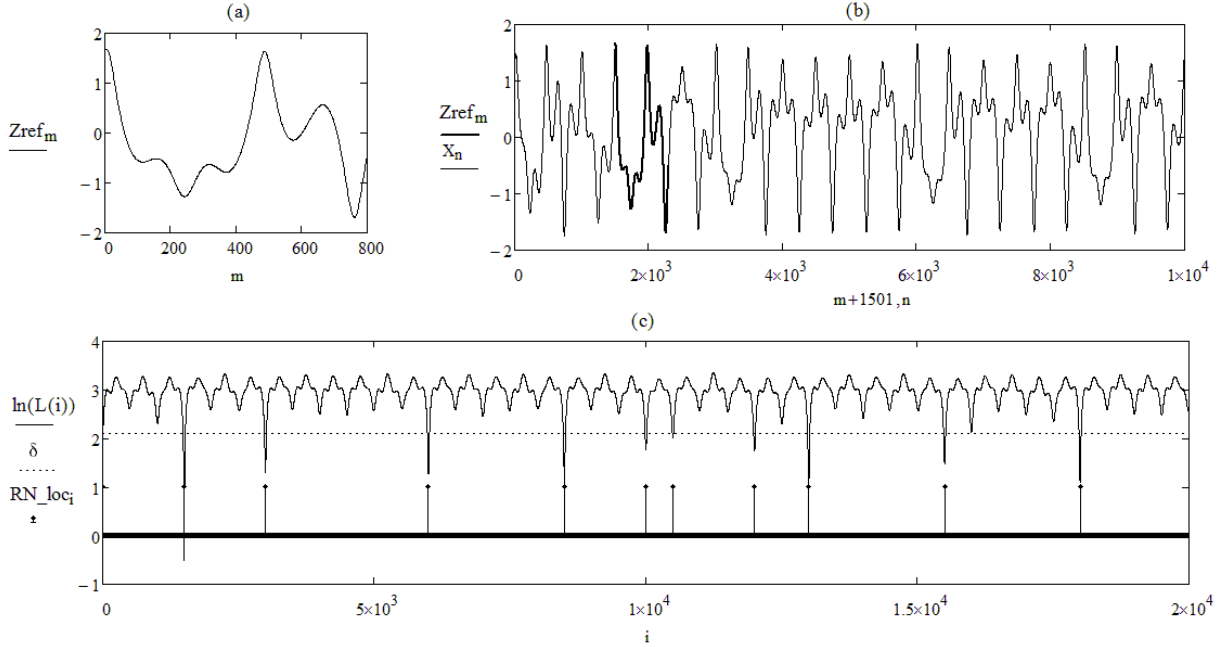}
    \caption{(a) $Z_{ref}$ is a pattern of length $800$ chosen from (b) a time-series generated by the Duffing Oscillator under Chaos, pattern (a) is highlighted in (b). (c) is $\ln$ of L(i)-- the distance of the neighbors with respect to $Z_{ref}$ (visualization of Eq.~3) and the stems of $RN\_loc$ plot correspond to the location of recurrent neighbors with respect to a chosen threshold $\delta$.} \label{Fig:timeseries}
  \end{center}
\end{figure}

\begin{figure}
\begin{centering}
\includegraphics [width=1 \textwidth]{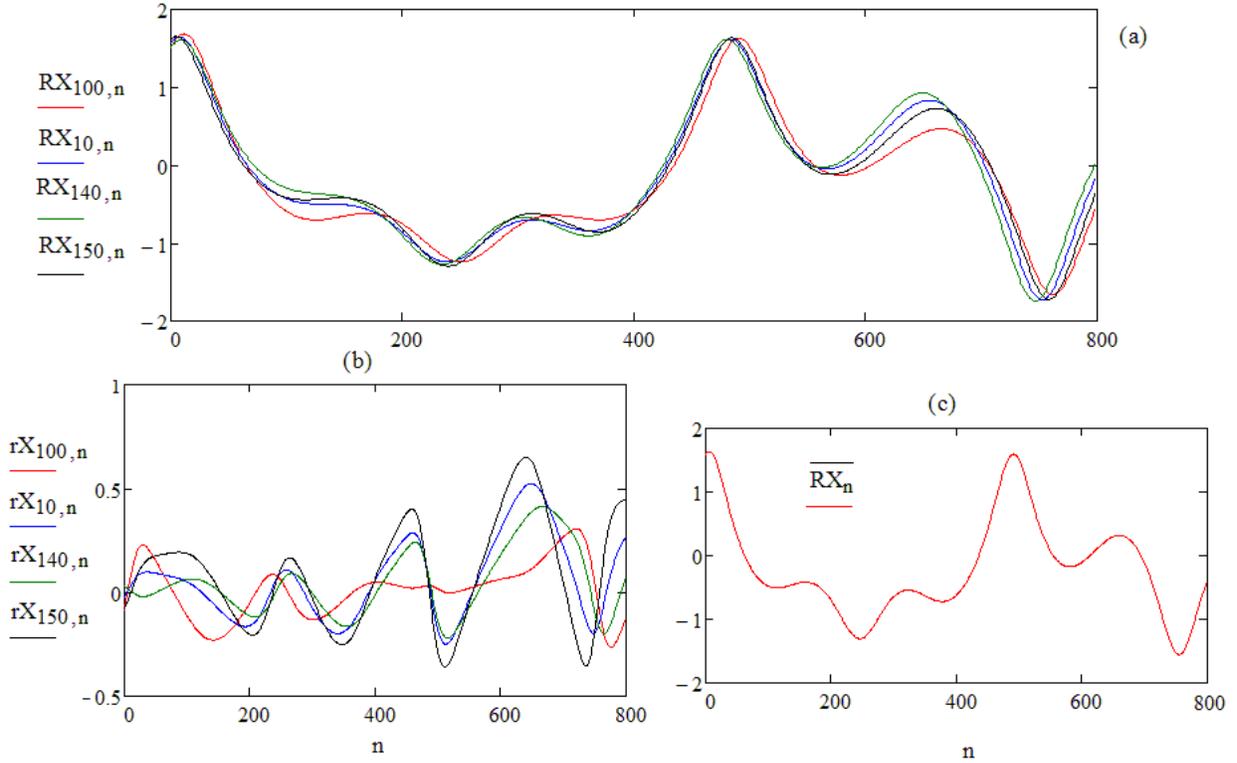}
\caption{(a) shows a few vectors in the recurrence neighborhood of $Z_{ref}$ in FIG.~\ref{Fig:timeseries}(a), (b) are the same vectors after translation to the origin, and (c) is the centroid $\overline{RX}$ of the recurrence neighborhood for chosen $Z_{ref}$. } \label{Fig:splitprofiles}       
\end{centering}
\end{figure}

A specific pattern $Z_{ref}$ is identified in the time-series generated by Duffing oscillator, to construct a recurrence neighborhood
for $Z_{ref}$ in $R^P$.  Figure~\ref{Fig:timeseries}(a) is the chosen pattern $Z_{ref}$ of length $P=800$ which is highlighted in ~\ref{Fig:timeseries}(b) the time-series generated by the Duffing Oscillator. Figure~\ref{Fig:timeseries}(c) shows  $\ln$ of L(i) 'the distance of $Z_{ref}$ with respect to all other vectors (visualization of Eq.~3)' and the location of recurrent neighbors with respect to a chosen threshold $\delta$ is shown by the stem plot $RN\_loc$. The neighbor who has minimum distance with $Z_{ref}$ in every recurrence cycle are chosen to belong to the recurrent neighborhood. Recurrence neighborhood for the chosen pattern $ Z_{ref}$ is identified in $R^P$ as explained in section III C.  Figure~\ref{Fig:splitprofiles}(a) shows a few vectors in the recurrence neighborhood of $Z_{ref}$ in Figure~\ref{Fig:timeseries}(a), and ~\ref{Fig:splitprofiles}(b) are the same vectors after translation to the origin, and (c) is the centroid $\overline{RX}$ of the recurrence neighborhood of chosen $Z_{ref}$.

\begin{figure}
\begin{centering}
\includegraphics [width=1 \textwidth]{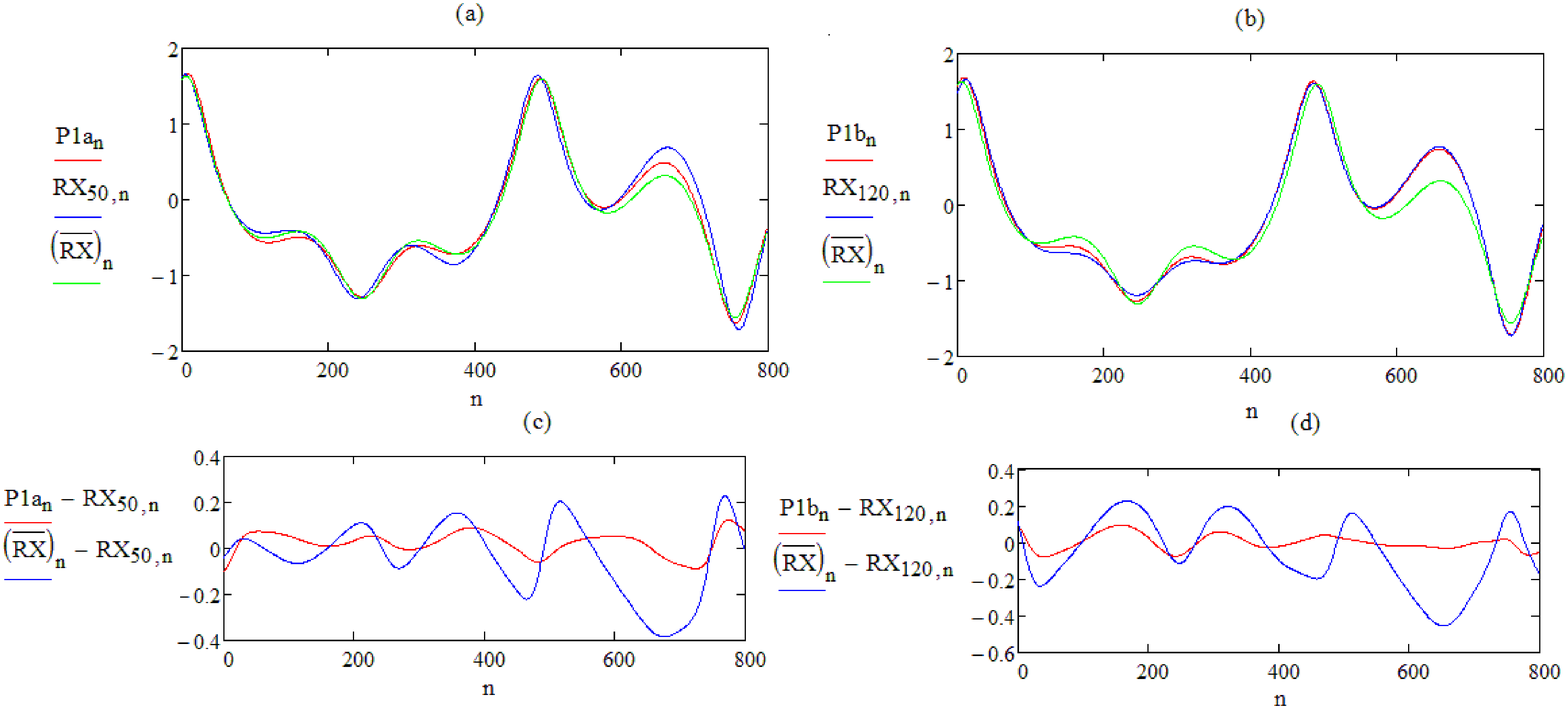}
\caption {$P1a, P1b$ are the Prediction scores for $50^{th}$ and $120^{th}$ vectors of $R^{800}$ neighborhood by affine maps. $\overline{RX}$ is the centroid fit. Predictions are shown by (a) and (b), corresponding the errors in prediction are given by (c) and (d).}
\label{Fig:4pred1}       
\end{centering}
\end{figure}

\begin{figure}
\begin{centering}
\includegraphics [width=1 \textwidth]{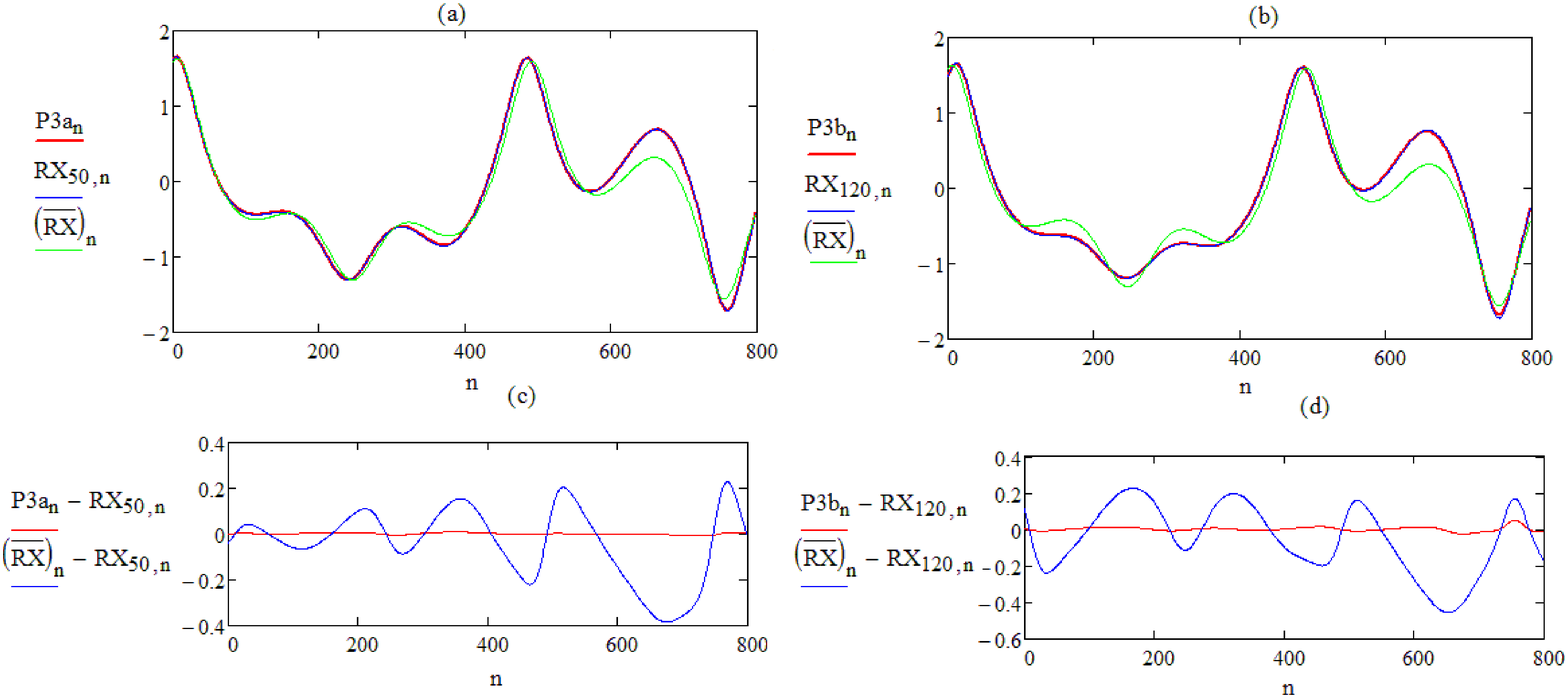}
\caption{$P3a, P3b$ are the improved prediction for $50^{th}$ and $120^{th}$ vectors of $R^{800}$ neighborhood by adding quadratic terms in the Atlas. $\overline{RX}$ is the centroid for $RX$. Predictions are shown by (a) and (b), corresponding the errors in prediction are given by (c) and (d). }
\label{Fig:4pred3}       
\end{centering}
\end{figure}

\begin{figure}
\begin{centering}
\includegraphics [width=1 \textwidth]{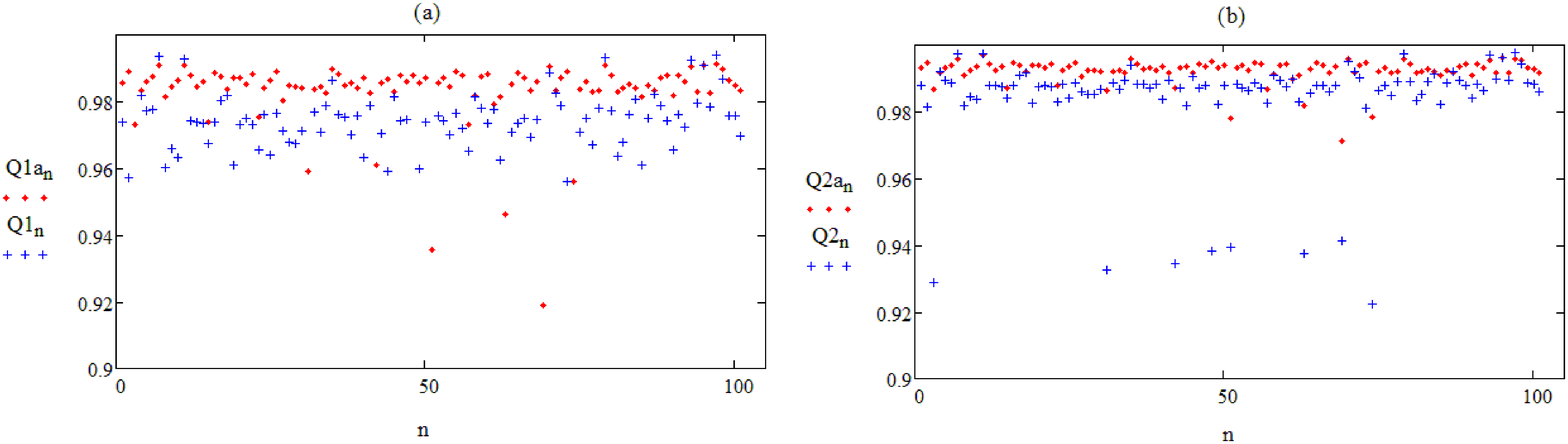}
\caption{Scores for 102 members of period-3 $\tau=1500$ equivalence class of vectors. Q1 and Q2 scores are for the affine predictions, Q1a and Q2a are improved scores by adding quadratic terms in Atlas.}
\label{Fig:4equclass_scores}       
\end{centering}
\end{figure}
\subsection{Parametrization for the recurrence neighborhood in $R^3$}

The mapping from high to low dimensional space ($R^{P} \rightarrow R^d$)  was done by the projective linear transformation of Eq.~4. The ARN model was tested for a choice of various pattern lengths $P$ in the range $500 \ldots 1800$, $d$ in the range $\{2,3,4 \}$ and $h_0 = \{500 \ldots 1000\}$. We infer from the numerical studies that the model is robust for these $P$ and $d$ values. For the graphs plotted in this section, the parameter values used were:  $h_0=500,  P=800$ and $d=3$.

\vspace{0.1in}
Results presented in this section are specific for the parametrization of the neighborhood in $R^3$ and the prescribed model aims to approximate the recurrence neighborhood in $R^{800}$ by a $3$-dimensional manifold. The neighborhood was approximated by a linear subspace and nonlinear manifold by minimizing the least squared errors using a standard and nonlinear Singular Value Decomposition~\cite{sajini2014}.

\vspace{0.1in}

\subsection{ Prediction Results by ARN Model}

Once the recurrence neighborhood, recurrence cycles, and equivalence classes are identified in $R^3$, prediction using conjugacy maps and evolution maps can be made as explained in section IV D. ARN models approximates the dynamics of the system by a few overlapping recurrence neighborhoods with specific maps for each of them. Results of prediction for the neighborhood vectors in $R^{800}$ using the parametrization maps in $R^3$ are presented here.
Predicted signals were compared with the original signals using two scores Q1~(based on mean squared errors) and Q2~(based on correlation). Scoring Functions are given in Appendix A.

\vspace{0.1in}
The Figures and scores given here are specific for the vectors that belong to the period-3 ($\tau=1500$) equivalence class. The period-3 equivalence class had $102$ members for the numerical simulation under trial.
Figure~\ref{Fig:4pred1} displays the prediction $P1a, P1b$ for two random members (indexed $50^{th},120^{th}$ in the recurrence neighborhood) that belong to the period-3 equivalence class: ~\ref{Fig:4pred1}(a) $P1a$ is the prediction for $50^{th}$ member using the affine map, $\overline{RX}$ is the centroid fit and $RX_{50}$ is the original vector, ~\ref{Fig:4pred1}(b) $P1b$ is the prediction for $120^{th}$ member using the affine map, $\overline{RX}$ is the centroid fit and $RX_{50}$ is the original vector, and ~\ref{Fig:4pred1}(c),(d) displays the errors in prediction for the affine fits and the mean fits with respect to the original vectors. Affine Predictions improve the mean fit. For the $50^{th}$ member Scores of Affine Prediction were $(Q1 = 0.9955, Q2 = 0.9985)$ and that of centroid fit were $(Q1 = 0.9604, Q2 = 0.9818)$. Scores of Prediction for the $120^{th}$ member were $(Q1 = 0.9975, Q2 = 0.9988)$ and that of centroid fit were $(Q1 = 0.9479, Q2 = 0.9780)$.

\vspace{0.1in}

Predictions can be further improved by fitting a nonlinear manifold by adding quadratic terms to the Atlas. Figure~\ref{Fig:4pred3} displays the improved predictions by adding nonlinear terms in the Atlas for the neighborhood.  Improved scores for $50^{th}$ member were $(Q1 = 0.99997, Q2 = 0.99999)$ and that for the $120^{th}$ member were $(Q1 = 0.9998, Q2 = 0.9999)$. Figure~\ref{Fig:4pred3} displays $P3a, P3b$ -- the Predictions, $\overline{RX}$ -- the centroid and the errors of the original signal with respect to the predicted signal and the centroid. The prediction scores for the affine map and their improved scores for all the members of a period-3 equivalence class is shown in  Figure~\ref{Fig:4equclass_scores}. Q1 and Q2 scores are for the affine prediction, Q1a and Q2a are improved scores by adding quadratic terms in Atlas.

\section{\label{sec:level1}Analysis of Real Data: Human Electrocardiogram (ECG) record }

This section considers a real-world data-- an electrocardiogram (ECG) of a human heart, to demonstrate the proposed ARN model.  A typical ECG signal has a repetitive structure in every cycle corresponding to various actions of the heart during a pumping cycle. Variation in beat-to-beat cycles of the ECG (Heart Rate Variability HRV) which can be seen as a variation in recurrent intervals is a typical characteristic of a healthy ECG measurement~\cite{HRvariabilityRef}. Since the signal exhibit both nonlinear and nonstationary features, both stochastic and nonlinear models are in use to capture essential aspects of the heart dynamics~\cite{porta2007,kantz_schrieber}.
\vspace{0.1in}

ARN model presented here has a fruitful point of view on ECG analysis. It approximates the dynamics of a system by a few overlapping recurrence neighborhoods with specific maps for each. For the particular case of ECG data, there is a possibility of deducing global structure from one or two small neighborhoods by fine-tuning the parameters of the ARN model, as demonstrated here.

\begin{figure}
\begin{centering}
\includegraphics [width=1 \textwidth]{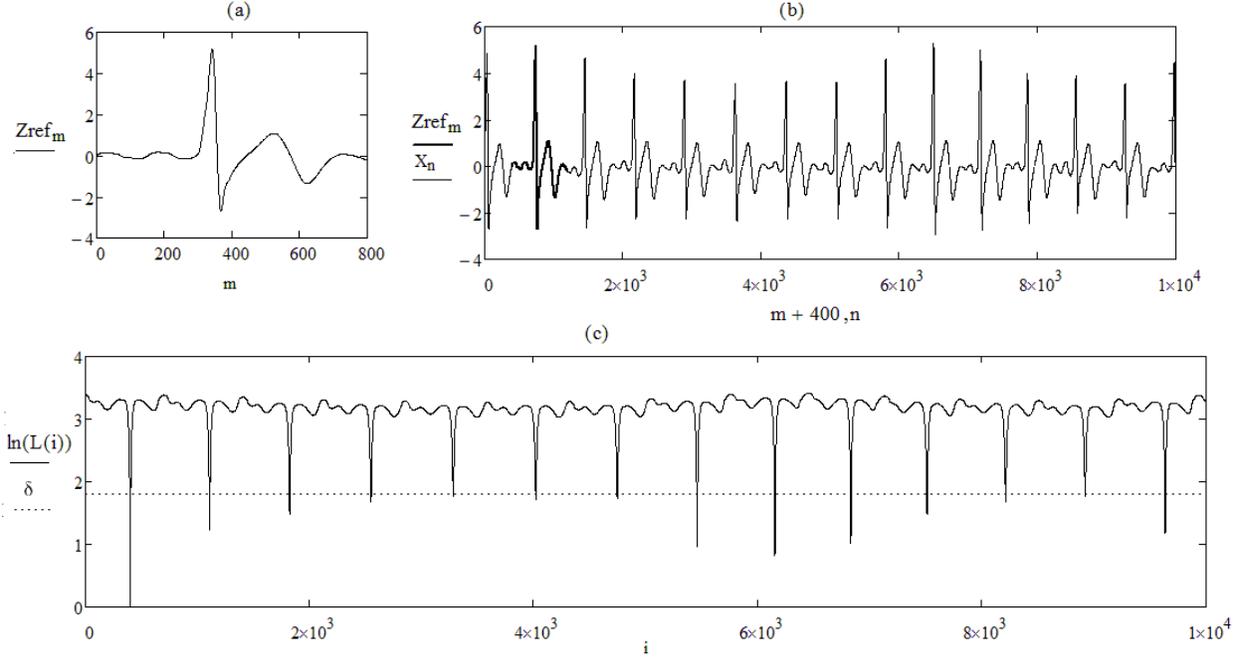}
\caption{(a) $Z_{ref}$ is a specific pattern of length $P=800$ that recur in an ECG data series shown in (b). (c) $L(i)$ indicates the distance of $Z_{ref}$ to every other vector $Z_i \in R^P$ with respect to $L2$ norm and it reflects that recurrence of the pattern (a) in the series.}
\label{Fig:5ECGtimeseries}       
\end{centering}
\end{figure}

\subsection{Identification of a Recurrence Neighborhood
for the ECG data}

\begin{figure}
\begin{centering}
\includegraphics [width=1 \textwidth]{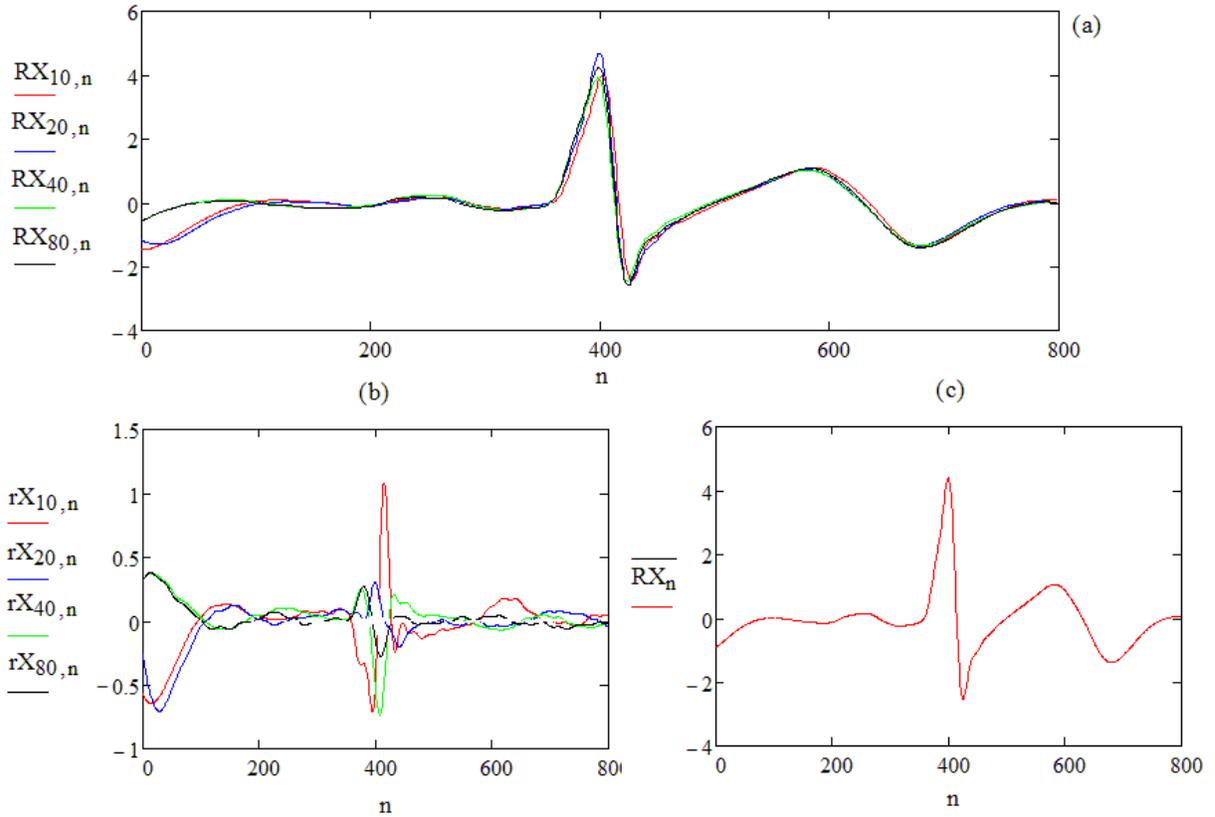}
\caption{(a) are a few vectors in the neighborhood of $Z_{ref}$ in $R^{P}$ before translation to the origin, (b) are the same vectors after translation to the origin, and (c) is the centroid $\overline{RX}$ of the neighborhood} \label{Fig:5splitprofiles}       
\end{centering}
\end{figure}

Figure~\ref{Fig:5ECGtimeseries}(a) is a specific pattern of an Electrocardiogram (ECG) record that represents a full cardiac cycle, which is highlighted in ~\ref{Fig:5ECGtimeseries}(b) of an ECG time-series. The time-series was recorded from a  human male aged 36 years, at a sampling frequency of  1kHz (data courtesy~\cite{data_nimhans}).
Figure~\ref{Fig:5ECGtimeseries}(a) is a complete cycle of heart rhythm (with standard patterns p-wave, QRS-complex and T-wave) that is chosen as the reference pattern $Z_{ref}$ of length $P=800$. Figure~\ref{Fig:5ECGtimeseries}(c) is the visualization $\ln$ of L(i)-- the distance of $Z_{ref}$ with respect to all other vectors, given by Eq.~3. Only the closest neighbor of $Z_{ref}$ (who has minimum distance with $Z_{ref}$) in every recurrence cycle belongs to the recurrent neighborhood.  Figure~\ref{Fig:5splitprofiles}(a) shows a few vectors in the recurrence neighborhood of $Z_{ref}$, (b) are the same vectors after translation to the origin and (c) is the centroid $\overline{RX}$ of the recurrence neighborhood. Following section VI E lists some algorithmic details about fine-tuning $P$ and $Z_{ref}$ for predicting full/ partial recurrence cycles.

If the chosen pattern $Z_{ref}$ contains precisely one cardiac cycle, then the recurrence intervals will match the beat-to-beat intervals or RR intervals of the ECG signal. Further, if the reference vector is selected such that it matches a QRS peak location, then the address of the recurrence neighbors will match with the QRS peak locations of the signal. Hence an algorithm to find recurrence neighbors can be tuned for finding QRS locations from ECG data.
According to the proposed model, if the recurrence timings are known for the ECG data, the rest of the dynamics is quite simple using the maps specific to the recurrence neighborhoods and its equivalence classes.

 \vspace{0.1in}

\subsection{Parametrization for the Recurrence Neighborhood in $R^d$}

The ECG signal under study was sampled at the rate of 1KHz and the finite time-series had 840 cardiac cycles, where its beat-to-beat intervals falling in a range of $646 -802$ ms.  The model was tested for a range of parameter values ($h_0 = \{ 500 \ldots 800\}$ and  $P = \{500 \ldots 800\}$ and $d =\{2,3,4\}$ and the results of  were found to be robust. For the graphs plotted in this section, the values used were:  $h_0=500,  P=800$, $d=3$ -- for the parametrization of the neighborhood in $R^3$. The neighborhood was approximated by a linear subspace and nonlinear manifold by minimizing the least squared errors~\cite{sajini2014}.

\vspace{0.1in}

\subsection{ Prediction Results by ARN model}

\begin{figure}
\begin{centering}
\includegraphics [width=1 \textwidth]{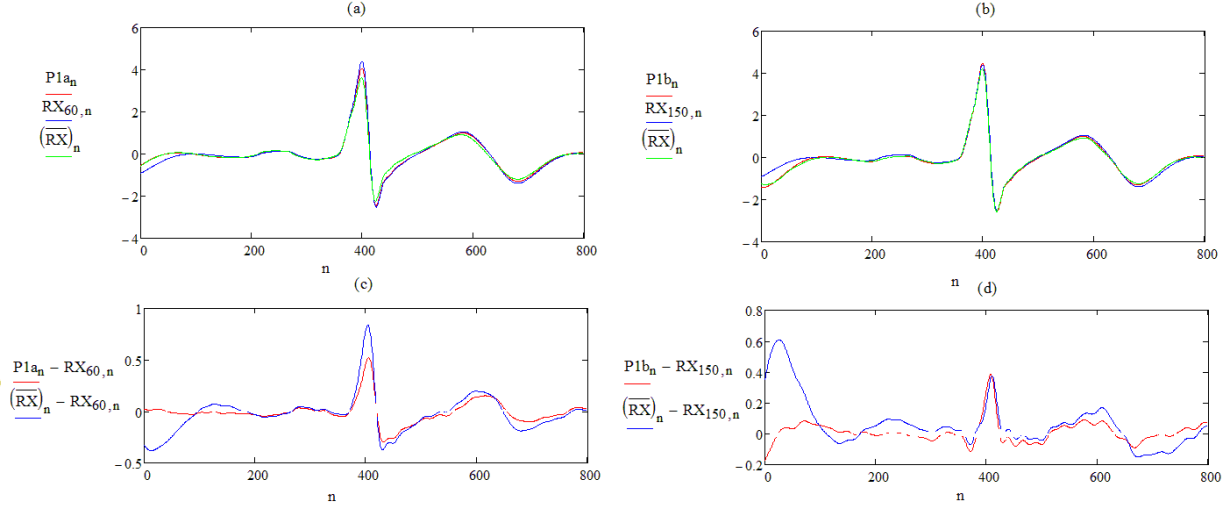}
\caption[Prediction from $R^3$ to $R^{800}$ based on standard SVD for two random vectors]{$P1a, P1b$ are prediction for $60^{th}$ and $150^{th}$ vectors of $R^{P}$ neighborhood $RX$. $\overline{RX}$ is the centroid for $RX$. Predictions are shown by (a) and (b), corresponding the errors in prediction are given by (c) and (d).}
\label{Fig:5pred1}       
\end{centering}
\end{figure}
\vspace{0.1in}
Predicted signals were compared with the original signals using two scores Q1(based on Mean Squared Errors) and Q2 (based on correlation).
Figure~\ref{Fig:5pred1} displays the prediction $P1a, P1b$ for two random members (indexed $60^{th},150^{th}$ in the recurrence neighborhood RX) that belong to the equivalence class corresponding to $\tau=695ms$. Figure~\ref{Fig:5pred1}(a) $P1a$ is the prediction for $60^{th}$ member using the affine map, $\overline{RX}$ is the centroid fit.  Similarly Figure~\ref{Fig:5pred1}(b) $P1b$ is the prediction for $150^{th}$ member using the affine map, $\overline{RX}$ is the centroid fit and $RX_{150}$.  $RX_{60}$ and $RX_{150}$ are the original vectors. Figure~\ref{Fig:5pred1}(c) and (d) displays the errors in prediction for the affine fits and the mean fits with respect to the original vectors. Predictions clearly improve the mean fit. Scores of Prediction for the $60^{th}$ member were  $(Q1 = 0.9830, Q2 = 0.9957)$ and that of centroid fit were $(Q1 = 0.9589, Q2 = 0.9906)$.  Scores of Prediction for the $150^{th}$ member were $(Q1 = 0.9945, Q2 = 0.9981)$ and that of centroid fit were $(Q1 = 0.9668, Q2 = 0.9855)$.
\vspace{0.1in}

\begin{figure}
\begin{centering}
\includegraphics [width=1 \textwidth]{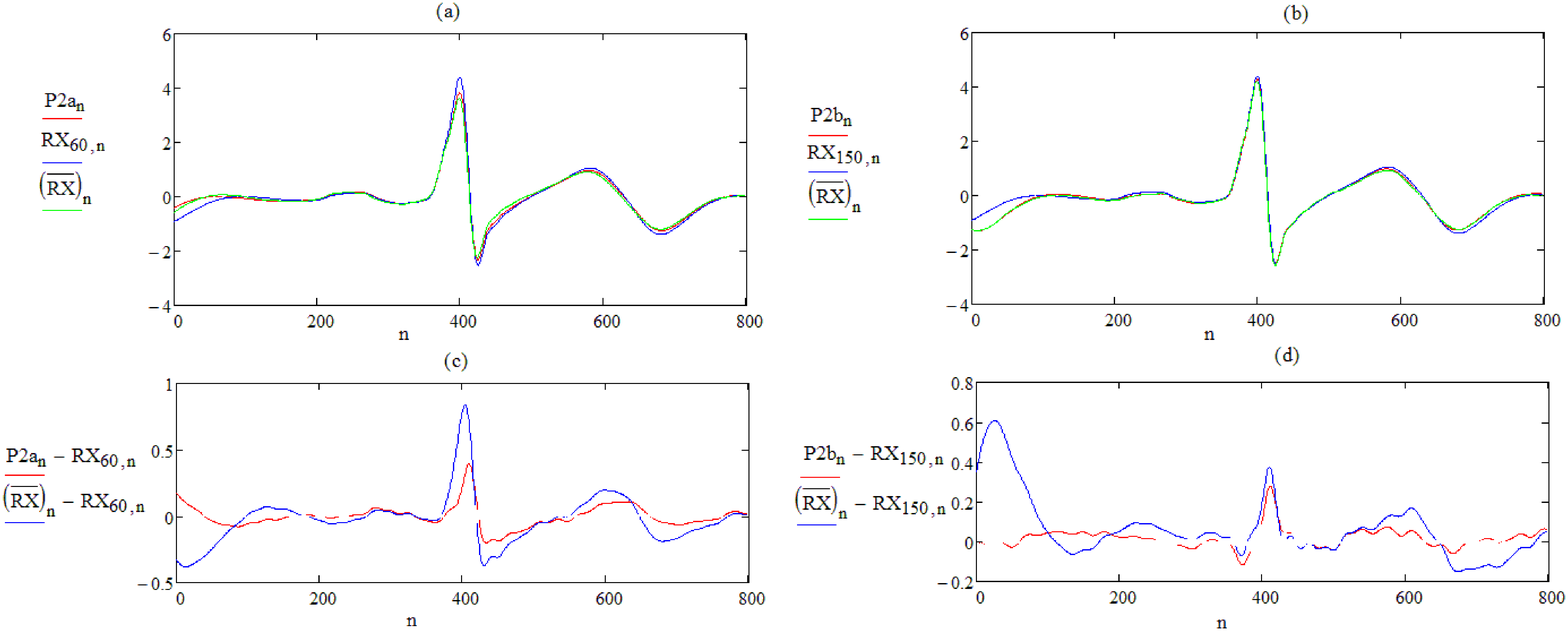}
\caption[Prediction from $R^3$ to $R^{800}$ based on nonlinear SVD for two random vectors]{$P2a, P2b$ are prediction for $60^{th}$ and $150^{th}$ vectors of $R^{800}$ neighborhood $RX$ using a nonlinear manifold. $\overline{RX}$ is the centroid for $RX$. Predictions are shown by (a) and (b), corresponding errors in prediction are given by (c) and (d).}
\label{Fig:5pred2}       
\end{centering}
\end{figure}

\begin{figure}
\begin{centering}
\includegraphics [width=1 \textwidth]{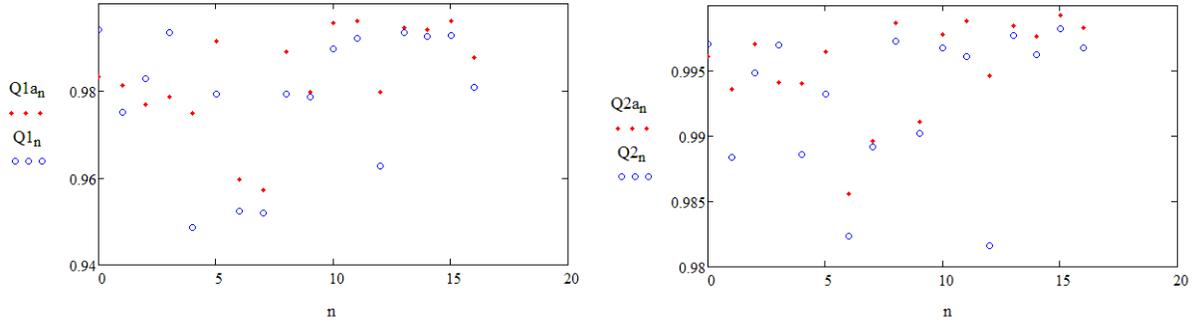}
\caption{Scores for 17 members of $\tau= 695 ms $ equivalence class of ECG data. Q1 and Q2 scores are for the affine map, Q1a and Q2a are improved scores after adding quadratic terms in Atlas}
\label{Fig:5equclass_scores}       
\end{centering}
\end{figure}

Predictions can be further improved by adding quadratic terms for the equation for the Atlas as we have seen in Section V. Figure~\ref{Fig:5pred2} displays the improved predictions by adding nonlinear terms in the Atlas for the neighborhood.  Improved Prediction scores for $60^{th}$ member were $(Q1 = 0.9897, Q2 = 0.9958)$ and that for the $150^{th}$ member were $(Q1 = 0.9969, Q2 = 0.9987)$. Figure~\ref{Fig:5pred2} (a), (b) displays $P3a,P3b$-- the nonlinear Predictions and $\overline{RX}$-- the centroid, (c) and (d) are corresponding the errors of prediction with respect to the the original signals. The prediction scores for the affine map, and the improved scores for $17$ the members of $\tau= 695 ms $ equivalence class (corresponding to RR interval $695 ms$) are shown in  Figure~\ref{Fig:5equclass_scores}. Q1 and Q2 scores are for the affine map, Q1a and Q2a are improved scores after adding quadratic terms in Atlas.

\subsection{Topological structure of ECG signals}
\begin{figure}
  \begin{center}
    \includegraphics [width= 0.8 \textwidth]{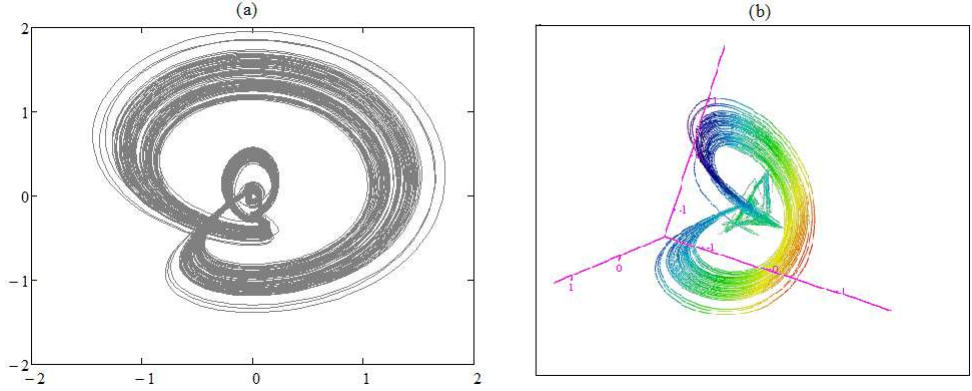}
    \caption{Global state-space reconstructed (a) in  $R^2$ and (b) $R^3$ and using a projection from the delay embedding space $R^{P}$}\label{Fig:R2SS}
  \end{center}
\end{figure}
Figure~\ref{Fig:R2SS} shows a reconstructed state-space of the ECG signal in  $R^2$ and $R^3$ after a projection from the delay-coordinate space $R^{P}$.  The property that the data lie in a ribbon-like manifold is used for prediction.  However, empirical estimates for correlation dimension from ECG data are known to be higher (in the range of \{9\ldots 15\} than $2$ or $3$~\cite{govindan1998}). However, the proposed model assumes that the dynamics heart can be seperated into two components: (i) the one that is responsible for the heart rate variability (ii) the effect of this variability on the pumping action of heart. The dynamics attributed to former `the variability in recurrences' is possibly higher due to its stochastic/nonlinear nature. The model assumes that if the higher order dynamics behind the variability in recurrences are taken apart, rest of the heart dynamics is simple (with a lower dimensional structure) that involves the switching of the equivalence classes corresponding to the recurrence cycles.
\vspace{0.1in}

In a deterministic dynamics perspective, points in a neighborhood of the state-space have similar future. An equivalence class specific to a particular $\tau$ recurrence cycle can be seen as a collection of neighbors that have similar future. Once equivalence classes specific to all recurrence cycles are known, global dynamics is reduced to a switching between the equivalence classes. For the ECG data under study,  an excellent prediction was obtained by using ARN model. Reconstruction of the data is done by stitching the embedding vectors appropriately at their corresponding recurrence locations.

\subsection{Specific algorithm to find Recurrence neighborhood for ECG data}

Find a pattern that contains one cardiac cycle with a single $QRS$ peak as $Z_{ref}$ whose length $P$ is approximately equal to an average recurrence cycle. Ideally, select a vector that has the QRS peak approximately towards the middle as $Z_{ref}$. Create a delay embedding of the data series in $R^P$ and find the similarity measure of other vectors $L$ with respect to $Z_{ref}$. Define a threshold $\delta$ such that all cardiac cycles are included, and find the neighbors that correspond to the local minima of $L$ to constitute the neighborhood RN. Record recurrence variability in an array RV-- the distance between two adjacent recurrences, and find the range of recurrence intervals-- from $RV_{min}$ to $RV_{max}$.
\vspace{0.1in}

To satisfy the condition that there should be exactly one QRS peak in $Z_{ref}$: either choose $P$ closer to the $RV_{max}$ (such that the QRS peak falls in the middle of $Z_{ref}$) or choose $P$ closer to the $RV_{min}$ (such that QRS peak falls in the beginning of $Z_{ref}$).  Reset the pattern $Z_{ref}$ and its length $P$ for fine-tuning. Once an optimal neighborhood is found, the prediction is straightforward using ARN model.
\vspace{0.1in}

\begin{figure}
\begin{centering}
\includegraphics [width=1 \textwidth]{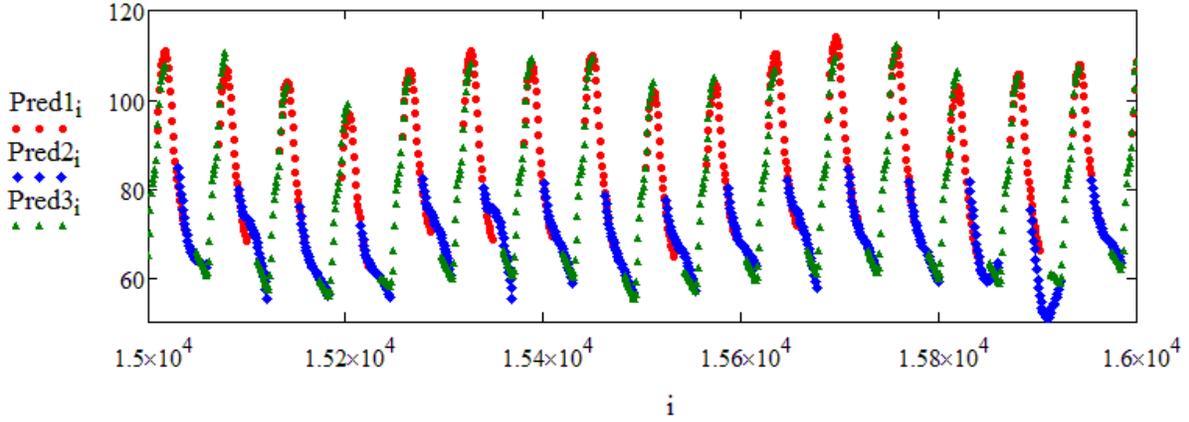}
\caption{Partial predictions Pred1 (i), Pred2 (ii) and Pred3 (iii) from 3 different recurrence neighborhoods for a blood pressure signal ABP from ECGII in \emph{c21.dat}~\cite{set_c_link} for $P=30$ (Choosing $P=110$ gave full predictions for this special case, and refer Table IV for the scores for \emph{c21.dat})}
\label{Fig:MPall3}       
\end{centering}
\end{figure}

If the  $P$ value is chosen much lesser than $RV_{min}$, one will have to consider multiple neighborhoods to cover the global reconstructed state-space. When $P$  is lesser than the optimal value,  vectors of the recurrent neighborhood in $R^P$ do not overlap the time-domain signal completely for the entire recurrent cycle resulting in a partial prediction of the signal. In that case, multiple recurrent neighborhoods can be used to predict the entire structure, by patching up the partial predictions. Figure~\ref{Fig:MPall3} shows partial predictions from 3 different recurrence neighborhoods for a blood pressure signal ABP from ECGII in \emph{c21.dat}~\cite{set_c_link} for $P=30$, whereas the optimal value  $P=110$ gave full predictions.

%

\section{Multichannel data-- Prediction across channels}
\begin{figure}
\begin{centering}
\includegraphics [width=1 \textwidth]{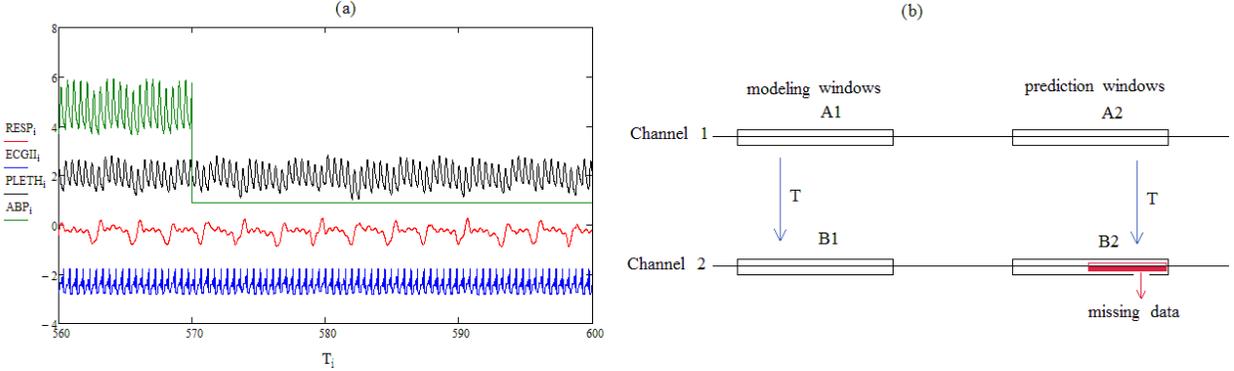}
\caption{(a) RESP, ECGII, PLETH, ABP signals~(scaled and shifted) in the last 40 seconds of a 10 minute record \emph{c21.dat}.  ABP signal missing in the last 30 seconds is denoted by a flat line. Prediction strategy is demonstrated in (b). Channel 2 contains the missing data segment for which a prediction is made using data available in channel 1. $A1, B1$ are the modeling windows across which a conjugacy map is identified, which is used for prediction for data in the window $B2$ from window $A2$.} \label{Fig:multichannels}
\end{centering}
\end{figure}

\begin{table}[!h]
\centering \caption{ECGII, ECGIII, ECGI Prediction Scores for data records in Set C~\cite{set_c_link}}
\vspace{0.1in}
\begin{tabular}{|l|l|l|l|}
\hline
Data record & Predicted signal & Score Q1 & Score Q2 \\
\hline
c08   & ECG II	& 0.9913	& 0.9965\\
\hline
c27   & ECG II	&0.9826	&0.9914\\
\hline
c33   &ECG II	&0.7318	&0.8623\\
\hline
c43   &ECG II	&0.9818	&0.9909\\
\hline
c45   &ECG II	&0.9987	&0.9994\\
\hline
c55   &ECG II	&0.9987	&0.9994\\
\hline
c70   &ECG II	&0.9982	&0.9991\\
\hline
c81   &ECG II	&0.9955	&0.9978\\
\hline
c86   &ECG II	&0.9922	&0.9961\\
\hline
c90   &ECG II	&0.9911	&0.9958\\
\hline
c98   &ECG II	&0.9782	&0.9898\\
\hline
c82   &ECGIII	&0.9923&	0.9968\\
\hline
c84   &ECGI	&0.0121	&0.6656\\
\hline
\end{tabular}
\label{tab:ECGIIpredictionscores}
\end{table}

\begin{table}[!h]
\centering \caption{ECGV Prediction Scores for data records in Set C~\cite{set_c_link}}
\vspace{0.1in}
\begin{tabular}{|l|l|l|l|}
\hline
Data record & Predicted signal & Score Q1 & Score Q2 \\
\hline
c02   &ECG V	&0.8857	&0.9429\\
\hline
c07   &ECG V	&0.9976	&0.9988\\
\hline
c18  &ECG V	&0.9976	&0.9988\\
\hline
c22   &ECG V	&0.8869	&0.9430\\
\hline
c23  & ECG V	&0.9602 &	0.9804\\
\hline
c30  &ECG V	&0.6921	&0.8516\\
\hline
c35   &ECG V	&0.9720	&0.9859\\
\hline
c49   &ECG V	&0.9948	&0.9974\\
\hline
c52   &ECG V	&0.4364	&0.7286\\
\hline
c57   &ECG V	&0.9831	&0.9922\\
\hline
c64   &ECG V	&0.9204	&0.9611\\
\hline
c75   &ECG V	&0.9245	&0.9626\\
\hline
c79   &ECG V	&0.9874	&0.9937\\
\hline
c89   &ECG V	&0.9763	&0.9882\\
\hline
c99   &ECG V	&0.8331	&0.9272\\
\hline
\end{tabular}
\label{tab:ECGVpredictionscores}
\end{table}

\begin{table}[!h]
\centering \caption{AVR Prediction Scores for data records in Set C~\cite{set_c_link}}
\vspace{0.1in}
\begin{tabular}{|l|l|l|l|}
\hline
Data record & Predicted signal & Score Q1 & Score Q2 \\
\hline
c09  & AVR	&0.9850&	0.9929\\
\hline
c10  & AVR	&0.9660&	0.9834\\
\hline
c29   &AVR	&0.9973	&0.9989\\
\hline
c38   &AVR	&0.2988	&0.6334\\
\hline
c58   &AVR	&0.9754	&0.9892\\
\hline
c71   &AVR	&0.8381	&0.9169\\
\hline
c74   &AVR	&0.9971	&0.9986\\
\hline
c78   &AVR	&0.9971	&0.9985\\
\hline
c80   &AVR	&0.9051	&0.9547\\
\hline
c88   &AVR	&0.9460	&0.9745\\
\hline
c92   &AVR	&0.9736	&0.9868\\
\hline
\end{tabular}
\label{tab:AVRpredictionscores}
\end{table}

\begin{table}[!h]
\centering \caption{ABP prediction scores for data records in Set C~\cite{set_c_link}}
\vspace{0.1in}
\begin{tabular}{|l|l|l|l|}
\hline
Data record & Predicted Signal & Score Q1 & Score Q2 \\
\hline
c01   &ABP	 &0.9229	 &0.9776 \\
\hline
c04    &ABP	 &0.9889	 &0.9951\\
\hline
c14    &ABP	 &0.7590	 &0.8837\\
\hline
c19    &ABP	 &0.9034	 &0.9723\\
\hline
c20    &ART	 &0.9914	 &0.9958\\
\hline
c21    &ABP	 &0.9724	 &0.9950\\
\hline
c31    &ART	 &0.9691	 &0.9895\\
\hline
c34    &ABP	 &0.9881	 &0.9956\\
\hline
c36    &ABP	 &0.7650	 &0.8827\\
\hline
c37    &ABP	 &0.9869	 &0.9945\\
\hline
c47    &ABP	 &0.9553	 &0.9779\\
\hline
c56    &ABP	 &0.9507	 &0.9952\\
\hline
c66    &ABP	 &0.9320	 &0.9680\\
\hline
c68    &ABP	 &0.9965	 &0.9984\\
\hline
c73    &ABP	 &0.8399	 &0.9172\\
\hline
c76    &ABP	 &0.9833	 &0.9920\\
\hline
c93    &ABP	 &0.9751	 &0.9875\\
\hline
c95    &ABP	 &0.9431	 &0.9803\\
\hline
\end{tabular}
\label{tab:BPpredictionscores}
\end{table}

This section explores capabilities of ARN model to analyze an extensive data set of multichannel recordings of physiological signals provided by Physionet~\cite{fullset_link}.  It contains three sets $A, B, C$ each containing $100$ multichannel cardiovascular recordings, which are $10$ minute in length, sampled at the rate of $125$ Hz. Signals in each of the multichannel recording vary across the records, and they include various ECG channels, blood pressure, respiration outputs. For a typical multichannel recording of length 10 minutes, the last $30$ seconds of a randomly chosen channel is replaced by a gap (a flat line signal) as shown in Figure~\ref{Fig:multichannels}(a). The goal of our study was to reconstruct the missing signal in each record using the information available in other channels using the ARN model and compare the efficiency of the predictions~(with respect to the originals that is available~\cite{set_c_link}).
\vspace{0.1in}

Predictions based on ARN model are based on the assumption that if two signals in the same multichannel recording exhibit the property of recurrence, there is a possibility of topological conjugacy across their respective recurrent neighborhoods. It can be assumed based on the following information. Events that happen in a multichannel record are related as they are simultaneous measurements of the same cardiovascular system. We can infer from the records that the signals are synchronized in some generalized manner with respect to each other sharing the property of recurrence. Some of the signals (all ECG and BP channels) have identical recurrent cycles with respect to each other, whereas some signals are synchronized in a phase locked behavior (ECG Vs Central venous pressure CVP, Respiration RESP). Figure~\ref{Fig:multichannels}(b) illustrates the algorithm for the ARN model.  Channel 2 contains the missing data segment and data available in Channel 1 is used for the prediction. $A1, B1$ are the modeling windows across which a conjugacy map is identified, for the prediction of data in the window $B2$ using data in the window $A2$.
\vspace{0.1in}

We infer from the analysis that the scores of reconstruction are dependent on the quality of the signals that are available on the channels. If the data available in the modeling and prediction windows are noisy or clipped, the predictions too were affected.  The parameters of the model estimated based on the modeling windows were used for prediction.  Though the signals are nonstationary and nonlinear, placing the modeling and prediction windows closer in time, the model assumes that the parameters are invariant across the windows. For a prediction of data in a $30 s$ window, a data of length $30 s$ ahead was used for the model.  Neural network based methods are the best among the current successful methods due to the development of novel training strategies~\cite{hinton1, hinton2}, which can solve complex problems and give great predictions with intense training sessions~\cite{haykin2004,Rodrigues2010,moody2010}. ARN model proposed here used estimated parameters from the modeling window and the lack of training has drastically reduced the time of prediction. Other than the excellent prediction capacity for the signals, lack of training time is an advantage of the ARN method over neural network methods.
\vspace{0.1in}

Tables~\ref{tab:ECGIIpredictionscores}--~\ref{tab:BPpredictionscores} contain $(Q1,  Q2)$ scores for all ECG and Blood pressure signals in set~C~\cite{set_c_link}. We find the efficacy of ARN models for predictions as excellent based on the scores, and we infer that the model could effectively reflect the changes in the cardiovascular system across the channels. The scores for rest of the signals of set C \{PLETH, RESP, ICP, CVP\} were not as good as the \{ECG, BP\} scores when the recurrent cycles were calculated with respect to the ECG signals~\cite{Saps2015PhDThesis}. Reasons for weak predictions could be: either there is no direct connection between the channels, or the connection was highly non-linear, or the model developed was inefficient to extract the connection. Since RESP and CVP have identical recurrence cycles with respect to each other (that is different from ECG, BP recurrence cycles), there is a possibility for improving scores by finding recurrent neighborhoods corresponding to their actual recurrent cycles.

\section{Conclusions}

We introduced Atlas for Recurrence neighborhood~(ARN) Model-- a novel topological method for prediction and modeling for a nonlinear time-series that exhibits recurring patterns. The paper demonstrated the model using~(i) a data generated by a well studied dynamical system--the Duffing oscillator under chaos and (ii) a real-world ECG data of a human. The proposed ARN model approximates the global manifold of the reconstructed state-space by a few overlapping recurrence neighborhoods. The computational load which is in general inevitable in nonlinear analysis was reduced as the model exploited the redundancy structure of the delay embedding procedure along with the property of recurrence.  Since corruption of cardiovascular signals are very common in real-time monitoring, potentials of the proposed ARN model to perform a challenging cognitive task--`the prediction of gaps or loss of data based on the contextual information' were explored in the study reported here, using a set of $100$ multichannel cardiovascular recordings. 

\section{Acknowledgments}

Sajini thanks Amit Apte for many constructive criticisms on topological modeling methods that has enhanced her understanding of the field. She thanks Nithin Nagaraj for numerous helpful discussions on this paper. She is grateful to ICTS--TIFR for providing a peaceful and motivating working environment during her survival of a major setback on health. She thanks National Board of Higher Mathematics, Department of Atomic Energy (NBHM--DAE) for supporting her post-doctoral research work at ICTS--TIFR.

\section{Appendix A: Scoring functions for evaluating Predictions}

Two types of scoring functions Q1~(based on mean squared errors) and Q2~(based on correlation) are used for comparing the prediction with the original signal. If $x_i$ and $y_i$ are the original signal and predicted signal respectively, where $i = 1, 2,...N$, then the scoring functions are defined as follows. $Q1(x,y)= 1-\frac{MSE(x,y)}{VAR(x)}$ and $Q2(x,y)=corr(x,y)$; where $MSE(x,y) = \frac{1}{N} \sum_{i=1}^N{(y_i - x_i)^2}$ and $VAR(x) = \frac{1}{N} \sum_{i=1}^N{{x_i}^2}-\left[\frac{1}{N} \sum_{i=1}^N{{x_i}} \right]^2$.
Note that Q1=1, Q2=1 implies $100\%$ fit between the original and the predicted signal.

\section{References}

\end{document}